\documentclass[a4paper]{ESASPStyle4astroph} 
\usepackage{graphics}
\usepackage{amssymb} 
\usepackage[mathcal]{eucal}

\newcommand{\apj} {ApJ}
\newcommand{\mnras} {MNRAS}
\newcommand{\aap} {A\&A}

\newcommand{\eqn} [1] {
\begin{equation} 
#1 
\end{equation}}

\newcommand{\eqna} [1] {
\begin{eqnarray} 
#1 
\end{eqnarray}}

\title{Oscillation power across the HR diagram : 
sensitivity to the convection treatment}
\author{R. Samadi \inst{1,3} \and G. Houdek \inst{2} \and M.-J. Goupil \inst{3}\and Y. Lebreton  \inst{3} \and A. Baglin \inst{3} }

 \institute{Astronomy Unit, Queen Mary, University of London, Mile End Road, UK \and  Institute of Astronomy, University of Cambridge, Cambridge CB3 0HA, UK \and 
Observatoire de Paris-Meudon, 5 place J. Janssen, F-92195 Meudon, France}

\begin{document}

\maketitle

\begin{abstract}
Solar-like oscillations are stochastically excited by turbulent convection.
In this work we investigate changes in the acoustic oscillation power spectrum
of solar-type stars by varying the treatment of convection in the equilibrium 
structure and the properties of the stochastic excitation model.
We consider different stellar models computed with the standard mixing-length 
description by \cite*{Bohm58} and with a generalized formulation of the 
mixing-length approach by Gough (1976, 1977\nocite{Gough76,Gough77}).
We calculate the acoustic power generated by the turbulent convection
which is injected stochastically into the acoustic pulsation modes.

Large differences in the oscillation powers are obtained depending on the 
choice of the assumed convection formulation. We show that the high-quality
data Eddington will provide, will allow us to distinguish between theoretical 
predictions of acoustic power spectra obtained with different convection 
models.
\end{abstract}

\section{Introduction}

The  amplitudes of solar-like oscillations are defined by the balance between 
excitation and damping. 
The oscillations are excited stochastically by the acoustic noise generated
by the turbulent motion of the convective elements.
In Samadi \& Goupil (2001, Paper\,I hereafter\nocite{Samadi00I}) a theoretical formulation for the 
acoustic power injected into solar-like oscillations is proposed and which 
supplements previous theories. 
We refer the reader to \cite*{Samadi01} for a detailed summary and discussion 
on some recent unsolved problems. 

The excitation process depends on the assumed turbulence spectrum, as 
discussed, for example, by \cite*{Samadi00b} and \cite*{Samadi01}. It also 
depends crucially
on the convection model to compute the stratification of the
convectively unstable layers in the equilibrium model. The amount of 
energy injected into the oscillations depends strongly on the velocity of 
the convective elements. 

The main goal of this work is to asses changes in the oscillation power 
spectrum due to modifications of the convection treatment in the equilibrium 
model.
We consider two different formulations: the classical description of 
mixing-length theory by \cite*{Bohm58} and a nonlocal generalization of the 
mixing-length formulation by Gough (1976, 1977\nocite{Gough76,Gough77}).
Additionally we study the dependence of the oscillation power on the assumed
turbulence spectrum in the excitation model for both convection formulations.

The theoretical formulation in Paper\,I involves two free parameters.
These parameters are calibrated such as to reproduce for a solar model the 
observed acoustic power spectrum (Section~2).
In Section~3 we compute oscillation power spectra for several stellar models 
using the two convection formulations.

We conclude that Eddington's performance will allow us to 
distinguish between the two treatments of convection and consequently that
Eddington will provide further constraints on convection models.

\section{Power injected into solar-like oscillations}

\subsection{Theory of the stochastic excitation}

The acoustic power $P$ injected into the oscillations is defined 
($e.g.$, \cite{GMK94}) in terms of the mode damping rate $\eta$, the oscillation
mean-square amplitude $\langle A^2 \rangle$, the mode inertia $I$ and 
oscillation frequency $\omega$ as~: 
\eqna{
P(\omega) = \eta\;  {\langle A^2 \rangle}\;I\;\omega^2\,.
\label{eq:P_omega}
}

The mean-square amplitude is defined by both the excitation by turbulent 
convection and by the damping process. It can be written
as
\eqna{
\left < A^2 \right > & \propto & \eta^{-1}
\int_{0}^{M}{\rm d}m \, \rho \, w^3 \, \Lambda^4 \,\left
(\frac{{\rm d}\xi_{\rm r}} {{\rm d}r} \right )^2    \nonumber \\ & & \times \,  
\left \{   \mathcal{S}_{\rm R} (\omega,m) + \mathcal{R}^2(m) \, 
\mathcal{F}^2( \xi_{\rm r},m) \, \mathcal{S}_{\rm S}  (\omega,m)\right \}\,,
\label{eqn:A2}
} %
where $\displaystyle{\xi_{\rm r}}$ is the radial displacement eigenfunction, 
$\rho$ the density, $\Lambda$ is the mixing length,  $w$ the vertical 
rms velocity of the convective elements;  $\mathcal{F}^2(\xi_{\rm r},m)$ is a 
function which includes the second derivative of $\xi_{\rm r}$, and 
$\mathcal{R}^2(m)$ describes the ratio of the excitation by the 
entropy fluctuations to that by the Reynolds fluctuations. 
The source functions $\mathcal{S}_{\rm R}(\omega,m)$ and 
$\mathcal{S}_{\rm S}(\omega,m)$ 
describe the contributions from the Reynolds and entropy fluctuations,
respectively, arising from the smaller scales of the turbulent cascade.
Detailed expressions for $\left < A^2 \right >$,  
$\mathcal{S}_{\rm R}(\omega,m)$,  
$\mathcal{S}_{\rm S}(\omega,m)$,  $\mathcal{R}^2$ and  $\mathcal{F}^2$  
are given in Paper\,I.

The source functions include the turbulent kinetic energy spectrum $E(k)$,
and the turbulent spectrum of the entropy fluctuations $E_s(k)$
which can be related to $E(k)$ by a simple expression 
($e.g.,$ \cite{Samadi00II}, Paper\,II hereafter). Both 
source functions $\mathcal{S}_{\rm R}(\omega,m)$ and 
$\mathcal{S}_{\rm S}(\omega,m)$ are integrated first over all eddy 
wavenumbers $k$,  followed by an integration over the stellar mass $M$ to 
obtain the acoustic power $P$ (see Eq.\ref{eqn:A2}).

In the present work, $\rho$ and $w$ are obtained from the two
equilibrium models, computed with the aforementioned two convection
formulations. The eigenfunctions $\xi_{\rm r}$ and eigenfrequencies $\omega$ 
are obtained from two different oscillation codes
and the $k$-dependence of $E(k)$  is inferred from different 
observations of the solar granulation and from theory.

\subsection{The equilibrium models}

We consider two sets of stellar models:
the first set consists of complete models obtained with the CESAM evolutionary 
code in which the convective heat flux is computed according to the classical 
mixing-length theory by {B\"ohm-Vitense} (1958, C-MLT hereafter\nocite{Bohm58}). The momentum flux
(sometimes referred to as turbulent pressure) is neglected in this set of
models. The eigenfunctions are obtained from the 
adiabatic FILOU pulsation code by \cite*{Tran95}. The detailed input physics 
used in this set of models is described in Paper~II.

The second set consists of envelope models computed in the manner of
\cite*{Balmforth92a} and \cite*{Houdek99}. In these models convection is 
treated with the nonlocal mixing-length formulation by Gough (1976, 
G-MLT hereafter\nocite{Gough76}), which consistently includes the 
momentum flux (i.e., the $rr$-component of the Reynolds stress tensor). The 
eigenfunctions are obtained from the nonadiabatic pulsation code by 
\cite*{Balmforth92a}, which includes both the Lagrangian perturbations of 
the heat and momentum fluxes in the manner of \cite*{Gough77}.

For both model sets the mixing length $\Lambda = \alpha \, H_p$, where $H_p$ 
is the local pressure scale height and $\alpha$ the mixing-length parameter,
is calibrated first to a solar model to obtain the helioseismically determined
depth of the convection zone of $0.287$ solar radii (\cite{C-DGT91}).
%

\subsection{Models for stellar turbulence}

\begin{figure}
	\resizebox{\hsize}{!}{\includegraphics{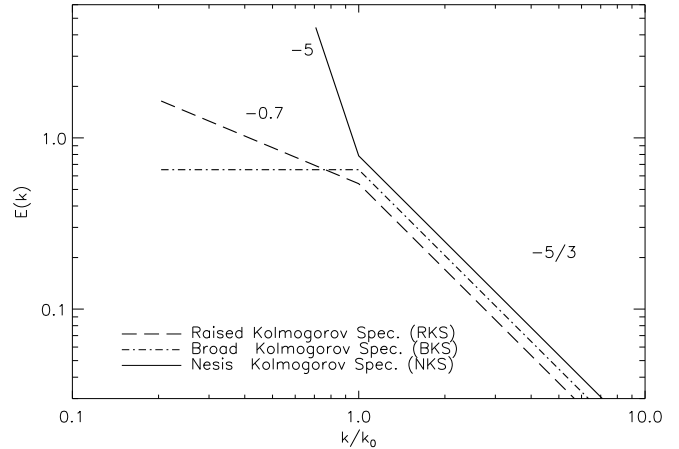}} 
	\caption{Kinetic turbulent spectra versus scaled wavenumber $k/k_0$.
		 }
	\label{fig:spc_cinetique}

        \end{figure}

Several turbulent spectra $E(k)$ were discussed in Paper~II:
the ``Nesis Kolmogorov Spectrum'' (NKS hereafter) and the
``Raised Kolmogorov Spectrum'' (RKS hereafter) were suggested from  
observations of the solar granulation by \cite*{Espagnet93} and
\cite*{Nesis93}. Here we also consider the ``Broad Kolmogorov Spectrum'' 
(BKS hereafter) by \cite*{Musielak94}. These spectra are depicted 
in Figure \ref{fig:spc_cinetique}.
For wavenumbers $k>k_0$, where $k_0$ is the smallest wavenumber of the 
classical Kolmogorov spectrum, all spectra follow the classical
Kolmogorov scaling law, $k^{-5/3}$. 
Only in the low-wavenumber range, $k < k_0$, where kinetic energy is injected
into the turbulent cascade, the spectra exhibit different scaling laws.
These spectra were considered by \cite*{Samadi00b} to compute acoustic power 
spectra for various solar-type stars and are also considered in this paper.

From observations of the solar granulation \cite*{Espagnet93} and 
\cite*{Nesis93} suggest various values for $k_0$.  Because of this 
ambiguity in the choice of $k_0$, we relate $k_0$
to the mixing length $\Lambda$ by $k_0=2\pi \, / \, (\beta \Lambda)$,
where $\beta$ is a free parameter of order unity (Paper\,I).
Another uncertainty in the formulation of the excitation model it the eddy 
correlation time scale. Because of our still poor understanding of turbulent
convection the eddy correlation time scale is not well defined and therefore
needs to be scaled, leading to an additional free parameter $\lambda$ of order
unity ($e.g.$, \cite{Balmforth92b}).
As it was demonstrated in Paper~II the oscillation power computed for the Sun 
depends crucially on the values of the free parameters $\lambda$ and $\beta$. 

\subsection{Calibration of the free parameters}

The mean square surface velocity $v_{\rm s}$ is related to the acoustic 
power $P(\omega)$ and the mode damping rate $\eta$ by the expression~: 
\eqn{
v_{\rm s}^2 = \xi_{\rm r}^2(r_{\rm s})  \;  P(\omega) \, / \, 2 \eta I\,,
\label{eqn:vs}
}
where  $r_{\rm s}$ is the radius at which the oscillations are observed. 

The acoustic power $P$ is computed for the two calibrated solar models 
using the aforementioned convection formulations.
For the velocity estimates $v_{\rm s}$, Eq.\ref{eqn:vs}, we assume 
the observed linewidths (damping rates) by \cite*{Libbrecht88}.  
The free parameters $\beta$ and $\lambda$ are calibrated for all turbulent 
spectra (see Section~2.3) and for both solar models to fit the estimated 
velocities $v_{\rm s}$ to the observations by \cite*{Libbrecht88}.

\begin{figure}
	\resizebox{\hsize}{!}{\includegraphics{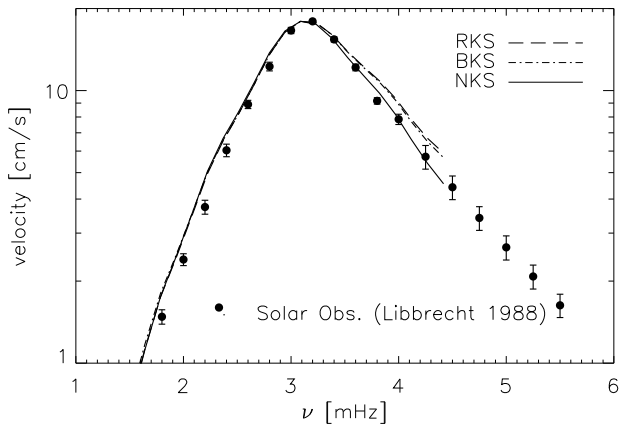}} 
	\caption{Computed surface velocities $v_{\rm s}$ (Eq.\ref{eqn:vs}) 
	assuming various turbulent spectra (see Fig.\,\ref{fig:spc_cinetique}) 
	and G-MLT for computing the turbulent fluxes. The free parameters 
	$\lambda$ and $\beta$ are calibrated to Libbrecht's (1988) velocity 
	and linewidth (damping rate) measurements.}
	\label{fig:VRSEP_sungh}
        \end{figure}

Figure~\ref{fig:VRSEP_sungh} displays the results for the solar model 
computed with G-MLT. The corresponding calibrated values of $\beta$ and 
$\lambda$ are listed in Table~\ref{tab:adjusted_parameters_G-MLT} 
(see Paper~II for the calibration results of the models computed with 
the C-MLT). 

At low frequencies the velocity $v_{\rm s}$ computed with G-MLT is in 
better agreement with the data than the results published in Paper~II 
obtained with the C-MLT.
At high frequencies the differences in $v_{\rm s}$ obtained with the 
various turbulent spectra are of similar small magnitude than the results 
of Paper~II which assumed the C-MLT.
Best agreement between theory and measurements is obtained with the NKS 
for both convection formulations.

\begin{table}
\caption{Calibrated values of the parameters $\beta$ and $\lambda$ for
the solar model computed with G-MLT.}
\label{tab:adjusted_parameters_G-MLT}
\begin{center}
\begin{tabular}{llll}  
spectrum&$\beta\lambda$&$\beta$&$\lambda$\cr
\hline
RKS & 0.6  & 1.96 & 0.31  \cr
BKS & 1.6  & 4.06 & 0.39  \cr
NKS & 2.6  & 3.11 & 0.83 \cr
\end{tabular}
\end{center}
\end{table}

\section{Scanning the HR diagram}

\subsection{Stellar models}

We consider various solar-type stars with masses between $1\,M_\odot$ and 
$2\,M_\odot$ 
in the vicinity of the main sequence. The model parameters are listed in
Table~\ref{tab:models_param}, which correspond to the models considered
previously by \cite*{Samadi00b}.
As for the solar models in Section~2 we consider two sets of stellar models:
the first set, computed with the C-MLT, is adopted from \cite*{Samadi00b}. 
The second set, computed with G-MLT, assumes the calibrated mixing length 
of the solar model discussed in Section~2.
The acoustical cut-off frequencies are very similar between the two sets 
of models and are displayed in Table~\ref{tab:models_param} for the first set.

\begin{table}[ht]
\begin{center}
\begin{tabular}{rrccccc}  
Models  & $L$ & $T_{\mathrm{\rm eff}}$ &  $M$ & Age & $ \nu_c$\\
 & [$L_\odot$] & [K] & [$M_\odot$] & [Gyr] & [mHz] \\
\hline 
$A$ & 12.1 & 6350 & 1.68 & 1.79 & 1.0  \\ 
$B$ &  9.0 & 6050 & 1.44 & 3.05 & 1.0 \\ 
$C$ &  6.6 & 6400 & 1.46 & 2.40 & 1.5 \\ 
$D$ &  3.7 & 5740 & 1.08 & 7.33 & 1.5 \\ 
$E$ & 3.5  & 6120 & 1.25 & 4.10 & 2.3 \\ 
$F$ & 2.6  & 6420 & 1.25 & 1.76 & 3.6  
\end{tabular}
\end{center}
\caption{Model parameters of solar-type stars. The model age and acoustical 
cut-off frequency $\nu_c$ are listed for the models obtained with the 
CESAM evolutionary code which assumes the C-MLT for computing the 
convective heat flux.
}
\label{tab:models_param}
\end{table}

\subsection{Dependence on convection models}

For all stellar models in both sets we assume the NKS and the calibrated 
values of $\beta$ and $\lambda$ quoted in \cite*{Samadi00b} and in 
Table~\ref{tab:adjusted_parameters_G-MLT}.

\begin{figure*}
\begin{center}
\resizebox{14cm}{!}{ \includegraphics{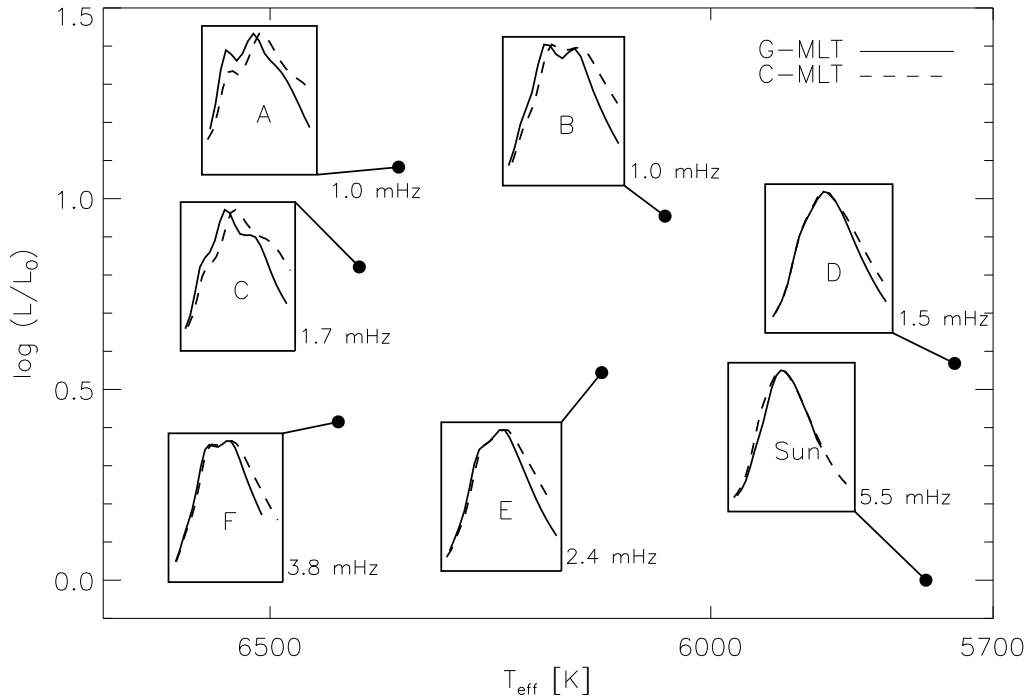}} 
\caption{Positions of models in the HR diagram (filled circles). For each 
model of the two sets, the computed oscillation power spectra assume the NKS 
and are illustrated in an associated box~: the solid curves correspond to models
computed with G-MLT and the dashed curves display the results obtained with the 
C-MLT. The plot range of the abscissa starts from zero and extends to the 
acoustic cut off frequency, indicated for each model. The power spectra are 
normalized with their maximum values.}
	\label{fig:pRSEPnkcs_stars2stars_HR}
\end{center}
        \end{figure*}

\begin{figure*}
	\resizebox{\hsize}{!}{\includegraphics{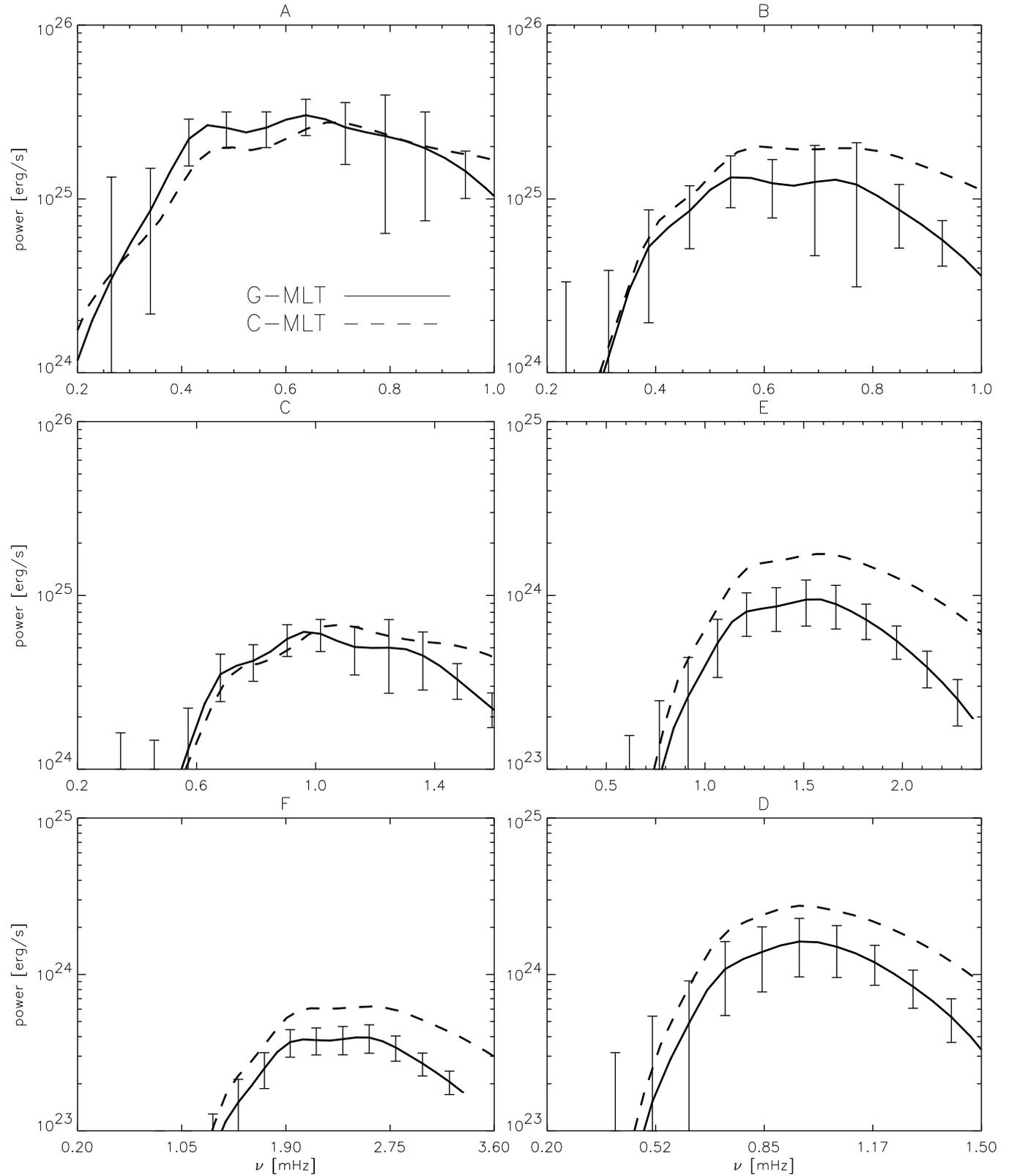}} 
	\caption{Oscillation power spectra computed for both sets of 
	stellar models. The NKS turbulence model is assumed. The solid curves 
	display the results for models computed with G-MLT and the dashed 
	curves show the results for models computed with the C-MLT.}
	\label{fig:pRSEP_CMLTvsGMLT}
        \end{figure*}

The position in the HR diagram of all stellar models and a qualitative
overview of their acoustic power spectra $P$ are depicted in
Fig.\,~ \ref{fig:pRSEPnkcs_stars2stars_HR}. Detailed results of $P$ are
shown in Fig.\,~\ref{fig:pRSEP_CMLTvsGMLT}.

\begin{figure*}
	\resizebox{\hsize}{!}{\includegraphics{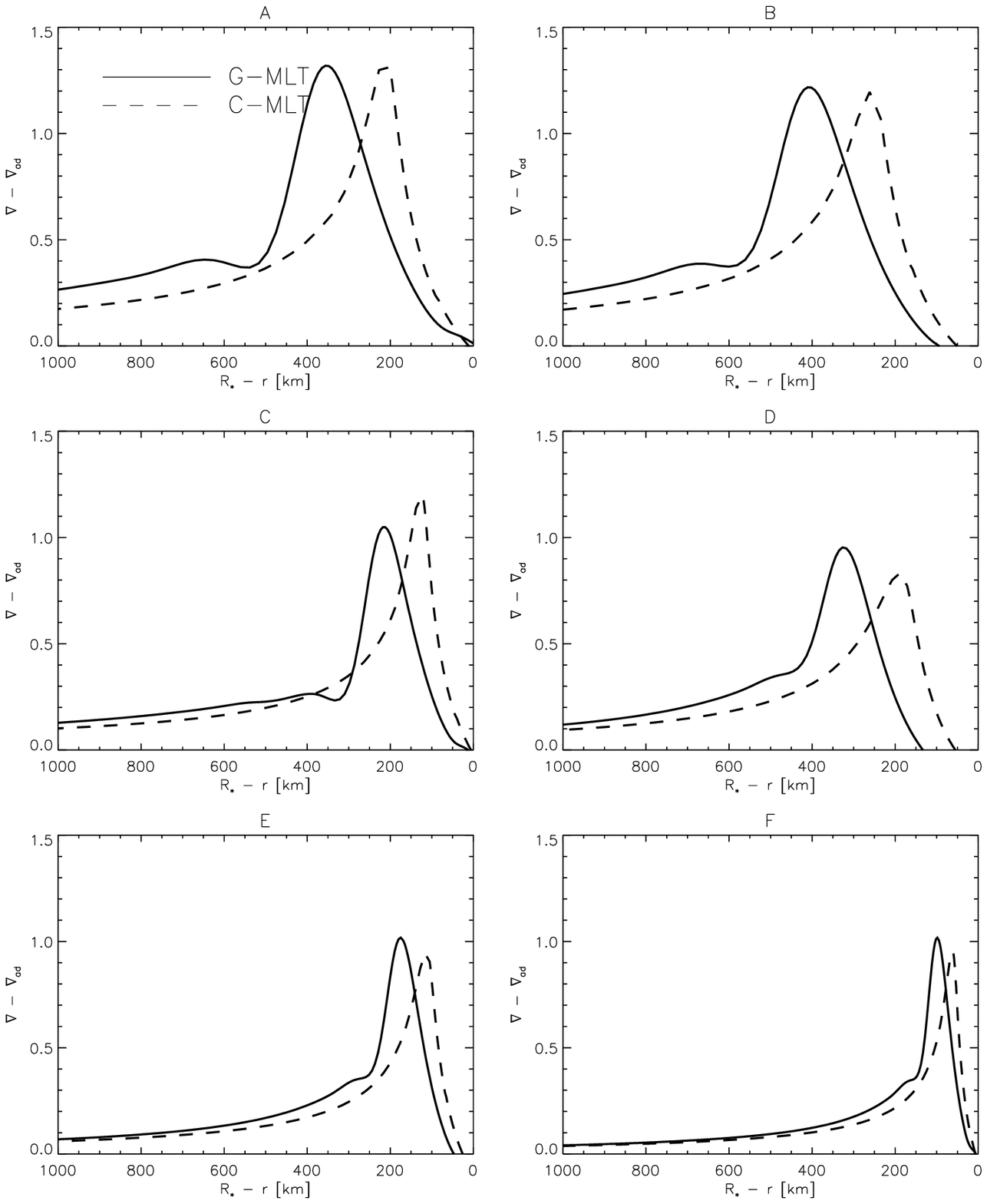}} 
	\caption{Depth dependence of the superadiabatic temperature gradient 
	$\nabla - \nabla_{ad}$ for the two sets of stellar models with surface
	radius $R_*$. The solid curves display the results for models computed 
	with G-MLT and the dashed curves are the results assuming the C-MLT.}
	\label{fig:cmp_gradT}
        \end{figure*}

At high frequencies the differences in $P$ between models computed with the 
C-MLT and G-MLT increase with increasing effective temperature $T_{\rm eff}$
and luminosity $L$.
As discussed in \cite*{Houdek96}, the nonlocal formulation (G-MLT) predicts
smaller temperature gradients in the upper superadiabatic region relative to 
the C-MLT. This means that convection is more efficient in models computed 
with G-MLT which leads to a different profile (depth-dependence) of the 
superadiabatic temperature gradient between the C-MLT and the G-MLT. The 
differences in the superadiabatic temperature gradient between the two model
sets increase with $L$ and $T_{\rm eff}$. These results are illustrated in 
Fig.\,~\ref{fig:cmp_gradT} which shows the superadiabatic temperature 
gradient $\nabla-\nabla_{ad}$ versus $R_* -r$, with $R_*$ being the radius 
of the star.
With increasing $L$ or $T_{\rm eff}$, the maximum of $\nabla - \nabla_{ad}$ 
is shifted more rapidly to deeper layers for the G-MLT models than for 
the C-MLT models. 
This leads to progressively larger differences in the convective velocities $w$
and in the shape of the eigenfunctions between the two sets of models. 
These differences in $w$ (note that $P$ depends crucially on $w$, see
Eq.\,\ref{eqn:A2}) and in the shape of the eigenfunctions are particular 
large in the superficial layers and result in a larger amount of acoustic 
power injected into high frequency modes for models computed with the C-MLT. 


There is an additional age effect: it can be seen from comparing the results 
of the models~E and F (see Fig.\,\ref{fig:pRSEP_CMLTvsGMLT}). Model E has the 
same mass than model F but is older. The increase of the maximum acoustic power 
with age is found to be larger for models computed with the C-MLT than for
the G-MLT models. 
%
%
%

\begin{figure*}
	\begin{center}
\resizebox{13.5cm}{!}{\includegraphics{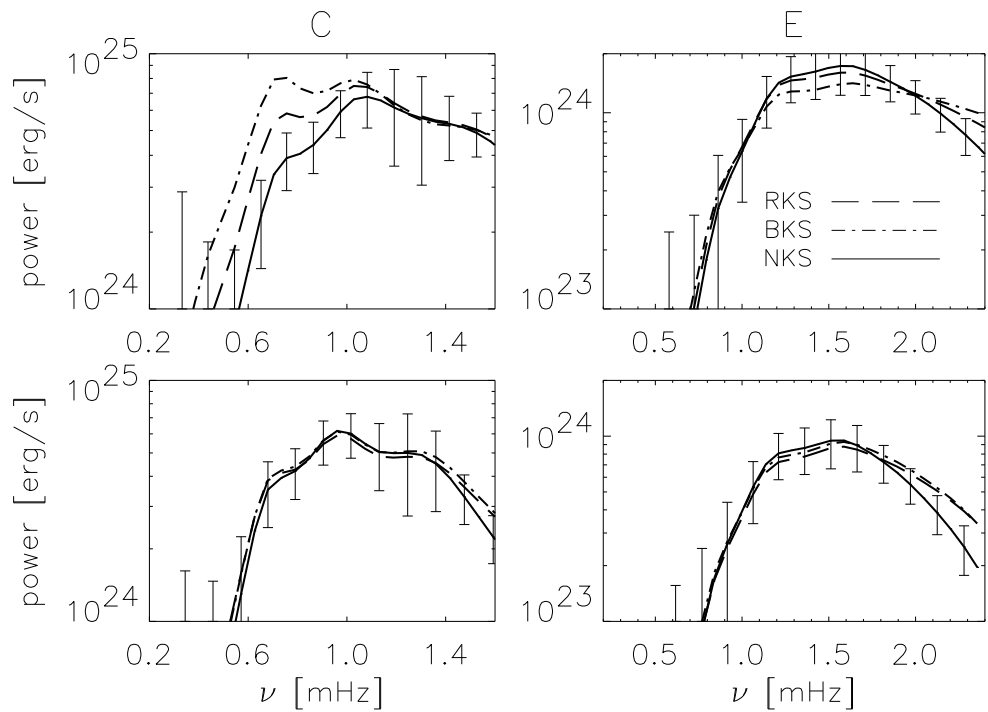}} 
\caption{Oscillation power spectra computed for the models C and E assuming 
the NKS (solid curves), the RKS (dashed curves) and the BKS 
(dot-dashed curves). The top panel displays the results for models computed
with the C-MLT and the bottom panel illustrates the results for models computed
with G-MLT. Vertical error bars $\Delta P$ are obtained from 
Eq.(\ref{eqn:DeltaP_P}) assuming an accuracy for $\eta/ 2\pi$ (resp. 
$\delta L/L$) of $0.3\,\mu$Hz (resp. $0.2$~ppm). 
}
	\label{fig:pRSEP_stars2stars_spc_CE}
 \end{center}
       \end{figure*}

\subsection{Dependence on turbulence models}

We compute $P$ for all the kinetic turbulent spectra discussed in
Section~2.3. The results are shown in Fig.\,\ref{fig:pRSEP_stars2stars_spc_CE} 
for the models C and E.
For the set of models computed with the C-MLT (top panels), the dependence 
of $P$ on the assumed turbulence spectra is more pronounced for the hotter 
model~C than for model~E.
This is not observed for the model set computed with G-MLT (bottom panels).

Models computed with the C-MLT exhibit a more pronounced
decrease of the depth of the surface convection zone with increasing 
$T_{\rm eff}$. This leads to a smaller extend of the excitation region 
for hotter models compared to the models computed with G-MLT. As discussed
in more detail by Samadi et~al. (2001b,c\nocite{Samadi00III,Samadi00b}) a shallower excitation
region results in a more pronounced frequency-dependence of $P$ on the
assumed turbulent spectra.
 
Model~C, obtained with G-MLT, exhibit a deeper excitation region and 
consequently the dependence of $P$ on the assumed turbulent spectra is smaller.
This property can be explained by the larger efficacy with which G-MLT 
transports the convective heat flux.

\subsection{Discussion}

One of Eddington's tasks will be the continuous observation of stellar 
luminosity oscillations over a time period of 30 days with a frequency
accuracy of $0.3\,\mu$Hz (\cite{Favata00}).
Furthermore the large telescope of Eddington (approximately ten times 
larger than that of the COROT mission) will lead to a noise level of
$\sim 0.2$~ppm for stars with a magnitude $m_v=6$ and for an observing period
of 30~days. For comparison, COROT will reach a noise level of $0.7$~ppm  
(\cite{Auvergne00}) for stars with a magnitude $m_v\simeq6$ and for a continuous
observing period of 5~days.

Will the high-quality observations of the Eddington mission be accurate 
enough to allow us to distinguish between the changes in the predicted
oscillation powers $P$ of Section~3.2 and Section~3.3 ?
In order to answer this question we estimate the expected accuracy 
of Eddington's measurements of $P$.

\cite*{Kjeldsen95} suggested a very simplified scaling law between
oscillation amplitude ratios $(\delta L/L)/v_{\rm s}$ and $T_{\rm eff}$:
\eqn{		
\delta L/ L  \propto v_{\rm s} \,  T_{\rm eff}^{-1/2}\,,
\label{eqn:dl_vs}
}
assuming adiabatic oscillations in a purely radiative star. It suggests that
the amplitude ratios scale inverse proportionally to the effective temperature
(but see also Houdek et~al., 1999, Fig.\,14, who found that the amplitude
ratios scale directly with $T_{\rm eff}$).

According to Eq.(\ref{eqn:vs}) and Eq.(\ref{eqn:dl_vs}) the relative error  
for $P$ can be expressed as
\eqn{
\frac{\Delta P}{P} = 2 \, \frac{\Delta \delta L}{\delta L} + 
\frac{\Delta \eta}{\eta}\,,
\label{eqn:DeltaP_P}
}
where $\delta L$ is obtained from Eqs.(\ref{eqn:vs}, \ref{eqn:dl_vs}). 
The damping rates $\eta$ are supplied from the nonadiabatic pulsation code of 
\cite*{Balmforth92a} assuming equilibrium models computed with G-MLT.

In Figure \ref{fig:pRSEP_stars2stars_spc_CE} and \ref{fig:pRSEP_CMLTvsGMLT}
the error bars $\Delta P$ are plotted according to Eq.(\ref{eqn:DeltaP_P})
assuming a noise level of $\Delta(\delta L)\sim 0.2$~ppm 
and $\Delta(\eta / 2 \pi)\sim 0.3 \, \mu$Hz.

We conclude that for hotter stars with a magnitude $m_v\leq6$, 
Eddington will provide data of sufficient accuracy which will allow us
to distinguish between the results obtained with the two considered 
convection formulations and consequently will provide further details 
on how to improve stellar convection models.

\begin{acknowledgements}
GH acknowledges the support by the UK Particle Physics and Astronomy 
Research Council.
\end{acknowledgements}


\begin{thebibliography}{19}
\expandafter\ifx\csname natexlab\endcsname\relax\def\natexlab#1{#1}\fi
\expandafter\ifx\csname url\endcsname\relax
  \def\url#1{{\tt #1}}\fi
\expandafter\ifx\csname urlprefix\endcsname\relax\def\urlprefix{URL }\fi

\bibitem[\protect \astroncite{{Auvergne} \& {the COROT Team}}{2000}]{Auvergne00}
{Auvergne} M., {the COROT Team}, 2000, In: The Third MONS Workshop :
Science Preparation and Target Selection, Aarhus University, 135--138

\bibitem[\protect \astroncite{Balmforth}{1992a}]{Balmforth92a}
{Balmforth} N.J., 1992a, \mnras, 255, 603

\bibitem[\protect \astroncite{{Balmforth}}{1992b}]{Balmforth92b}
{Balmforth} N.J., 1992b, \mnras, 255, 639

\bibitem[\protect \astroncite{{B\"ohm-Vitense}}{1958}]{Bohm58}
{B\"ohm - Vitense} E., 1958, Zeitschr. Astrophys., 46, 108

\bibitem[\protect \astroncite{Christensen-Dalsgaard et~al.}{1991}]{C-DGT91}
{Christensen-Dalsgaard} J., {Gough} D.O., {Thompson} M., 1991, \apj, 378, 413

\bibitem[\protect \astroncite{{Espagnet} et~al.}{1993}]{Espagnet93}
{Espagnet} O., {Muller} R., {Roudier} T., {Mein} N., 1993, \aap, 271, 589

\bibitem[\protect \astroncite{{Favata} et~al.}{2000}]{Favata00}
{Favata} F., {Roxburgh} I., {Christensen-Dalsgaard} J., 2000, In: The
  Third MONS Workshop : Science Preparation and Target Selection, 
  Aarhus University, 49--54

\bibitem[\protect \astroncite{{Goldreich} et~al.}{1994}]{GMK94}
{Goldreich} P., {Murray} N., {Kumar} P., 1994, \apj, 424, 466

\bibitem[\protect \astroncite{Gough}{1976}]{Gough76}
{Gough} D., 1976, In: {Spiegel} E., {Zahn} J.P. (eds.) Problems of stellar
  convection, Vol.~71 of Lecture notes in physics, 15, Springer Verlag

\bibitem[\protect \astroncite{{Gough}}{1977}]{Gough77}
{Gough} D.O., 1977, \apj, 214, 196

\bibitem[\protect \astroncite{{Houdek}}{1996}]{Houdek96}
{Houdek} G., 1996, Pulsation of Solar-type Stars, Ph.D. Thesis, 
       Institut f\"ur Astronomie, Wien

\bibitem[\protect \astroncite{Houdek et~al.}{1999}]{Houdek99}
{Houdek} G., {Balmforth} N.J., {Christensen-Dalsgaard} J., {Gough} D.O.,
  1999, \aap, 351, 582

\bibitem[\protect \astroncite{{Kjeldsen} \& {Bedding}}{1995}]{Kjeldsen95}
{Kjeldsen} H., {Bedding} T.R., 1995, \aap, 293, 87

\bibitem[\protect \astroncite{Libbrecht}{1988}]{Libbrecht88}
{Libbrecht} K.G., 1988, \apj, 334, 510

\bibitem[\protect \astroncite{Musielak et~al.}{1994}]{Musielak94}
{Musielak} Z.E., {Rosner} R., {Stein} R.F., {Ulmschneider} P., 1994, \apj,
  423, 474

\bibitem[\protect \astroncite{{Nesis} et~al.}{1993}]{Nesis93}
{Nesis} A., {Hanslmeier} A., {Hammer} R., et~al., 1993, \aap, 279, 599

\bibitem[\protect \astroncite{Samadi}{2001}]{Samadi01}
{Samadi} R., 2001, In: SF2A-2001: Semaine de l'Astrophysique Francaise,  E148

\bibitem[\protect \astroncite{{Samadi} \& {Goupil}}{2001}]{Samadi00I}
{Samadi} R., {Goupil} M., 2001, \aap, 370, 136

\bibitem[\protect \astroncite{{Samadi} et~al.}{2001a}]{Samadi00II}
{Samadi} R., {Goupil} M., {Lebreton} Y., 2001a, \aap, 370, 147

\bibitem[\protect \astroncite{{Samadi} et~al.}{2001b}]{Samadi00b}
{Samadi} R., {Goupil} M.J., {Lebreton} Y., {Baglin} A.,
  2001{\natexlab{b}}, In: {Wilson} A. (ed.) Proceedings of the SOHO 10/GONG
  2000 Workshop, 'Helio-and Asteroseismology at the Dawn of the Millennium',
  vol. SP-464 (astro-ph/0101129), 451--456, ESA Publications Division

\bibitem[\protect \astroncite{{Samadi} et~al.}{2001c}]{Samadi00III}
{Samadi} R., {Houdek} G., {Goupil} M.J., {Lebreton} Y., 2001{\natexlab{c}},
  to be submitted to \aap

\bibitem[\protect \astroncite{{Tran Minh} \& {Leon}}{1995}]{Tran95}
{Tran Minh} F., {Leon} L., 1995, In: Physical Process in Astrophysics, 219

\end{thebibliography}
\end{document}